\documentclass[prb,twocolumn,showpacs,amsmath,amssymb]{revtex4}

\usepackage{graphicx}
\usepackage{dcolumn}
\usepackage{bm}
\usepackage{epsfig}

\begin{document}

\title{Mesoscopics in Spintronics: Quantum Interference Effects in
Spin-Polarized Electron Transport}

\author{Branislav K. Nikoli\' c}
\altaffiliation[Present address: ]{Department of Physics and Astronomy, University of Delaware, Newark, DE 19716-2570}
\author{J. K. Freericks}
\affiliation{Department of Physics, Georgetown University,
Washington, DC 20057-0995}

\begin{abstract}
We generalize a Landauer-type formula, using a real$\otimes$spin-space Green function technique, 
to treat spin-dependent transport in quantum-coherent conductors 
attached to two ferromagnetic contacts. The formalism is employed
to study the properties of components of an exact zero-temperature 
conductance matrix ${\bf G}$, as well as their mesoscopic fluctuations, 
describing injection and detection of a spin-polarized current in a 
two-dimensional system where electrons exhibit an interplay between 
Rashba spin-orbit (SO) coupling and phase-coherent propagation through a 
disordered medium. Strong Rashba coupling leads to a dramatic reduction of 
localization effects on the conductances and their fluctuations, whose features 
depend on the spin-polarization of injected electrons. In the limit of weak
Rashba interaction antilocalization vanishes (i.e., the sum of the matrix elements 
of ${\bf G}$ is almost independent of the SO coupling), but the partial 
spin-resolved conductances can still be non-zero. Besides spin-resolved conductance 
fluctuations and antilocalization, unusual  quantum interference effects are revealed in 
this system leading to a negative difference between the partial conductances for a 
parallel and  an antiparallel orientation of the contact magnetization, in a range of disorder 
strengths and for a particular spin-polarization of incoming electron with respect to 
the direction of Rashba electric field.
\end{abstract}

\pacs{72.25.-b, 72.25.Rb., 73.23.-b, 85.75.Hh.}

\maketitle

\section{Introduction}\label{sec:intro}

Due to its fundamentally quantum nature, electron spin has very limited 
choice of interactions through which it can couple to the environment. Thus, its 
polarization can survive various forces in metals or semiconductors, while charge carriers 
undergo many scattering events, for long enough time to allow for envisioned quantum 
technologies that manipulate spin, such as spintronics~\cite{spintronics,datta90} 
or solid-state quantum computing with spin-qubits.~\cite{qc} Spintronic engineering of 
spin-polarized currents, combined with conventional electronics that manipulate electron 
charge,  offers an exciting prospect to assemble information processing, storage, and 
communication on the same chip. Recent experimental advances have provided an impetus 
to study spin-polarized transport in two-dimensional electron systems, posing theoretical 
challenges~\cite{seba} where the ubiquitous ``factor of two'' for spin degeneracy in usual 
transport formulas has to be replaced by a more involved analysis of the interplay between 
orbital and spin degrees of freedom of the electrons.
\begin{figure}
\centerline{ \psfig{file=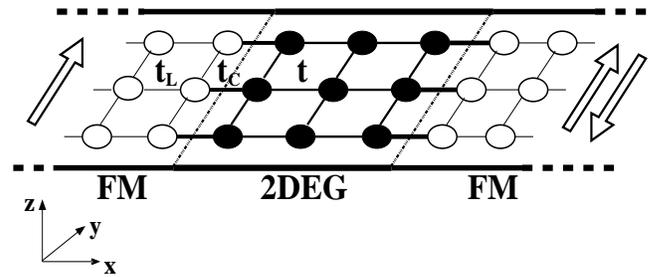,height=1.4in,width=3.4in,angle=0}
} \caption{Graphical depiction of our lattice model for a generic 
two-probe quantum spintronic FM-Sm-FM device: two ferromagnetic contacts (FM), that act as a spin 
injector (left) and a spin detector (right) are attached to a two-dimensional 
electron gas (e.g., in InAs), which is confined to the $xy$-plane by a Rashba 
electric field along the $z$-axis. The white arrows denote one possible magnetization 
configuration for the spin-resolved measurements (e.g., spin-up along $y$-axis is injected 
and both spins are collected). Each site hosts a single $s$-orbital which hops 
to four (or fewer for surface atoms) nearest neighbors. The hopping matrix element 
is $t$ (within the sample), $t_{\rm L}$ (within the FM leads), and $t_{\rm C}$ (coupling of 
the sample to the leads). The leads are semi-infinite and connected smoothly at $\pm \infty$ 
to macroscopic reservoirs biased by the chemical potential difference $\mu_L-\mu_R=eV$.}
\label{fig:setup}
\end{figure}

While quantum-coherence of spin is essential for spintronic qubits, as well as some of 
the proposed spintronic devices for classical information processing,~\cite{datta90} modeling of 
the orbital kinetics of electrons that carry spin has mostly followed the traditional route 
of device engineering in charge-electronics: the charge transport (ballistic or diffusive) 
is described by a semiclassical Boltzmann formalism  treating electrons and holes as classical 
particles where quantum effects enter only through their kinetic energy (being determined by the semiconductor 
band structure).~\cite{semi,semi1} Although many applications require devices operating at room 
temperature (where phase-coherence of orbital degrees of freedom is usually washed out), 
from the fundamental transport physics point of view, it is intriguing to explore quantum  
interference effects involving both the charge and spin of the electrons in nanoscale samples 
at very low temperatures $\lesssim 1$K---a playground of mesoscopic physics over the 
past two decades.~\cite{mesophys} Moreover, better understanding of the role of conserved 
quantum coherence of both wave functions and spin states in the conductions process is crucial for 
the devices that involve spin-polarized resonant tunneling.~\cite{yuasa} Recent theoretical efforts 
have been exploring mesoscopic aspects of spintronics by extending standard techniques (such as, the  
Landauer-B\" uttiker scattering approach to quantum transport of {\em spinless} particles in 
finite-size systems~\cite{datta,carlo_rmt}) to describe transport in the presence of spin-dependent 
interactions by taking into account quantum-coherent dynamics of both orbital and spin degrees of 
freedom.~\cite{seba,mireles} Even though it is possible to add local spin-dependent effects into the 
scattering matrix directly, such extensions become highly non-trivial in the most general case where 
spin effects occur all over the system, as in the  presence of SO interactions.~\cite{seba,feve} 

A particularly important interaction for semiconductor spintronics 
is the Rashba type of spin-orbit coupling,~\cite{rashba} which is
different from the more familiar impurity induced and position
dependent one in metals.~\cite{larkin} It arises from the
asymmetry along the $z$-axis of the confining quantum well
electric potential (generating a magnetic field 
in the electron rest frame) that creates a two-dimensional 
(2D) electron gas ($xy$-plane) on a narrow-gap semiconductor surface. Since the
Rashba coupling can be tuned by an external gate
electrode,~\cite{nitta} it is envisaged as a tool to control the
precession of the electron spin in the Datta-Das proposal
for a field-effect spin  transistor.~\cite{datta90} Thus, in such
spin-interference devices intrinsic magnetic degrees  of freedom
can be controlled by electrical means. The spin-FET embodies a
paradigm of a spintronic two-terminal device, such as the one
shown in Fig.~\ref{fig:setup}: {\bf (i)} at the source
(injector), information is stored as spins in a particular 
orientation (spin-up $\uparrow$ or spin-down $\downarrow$); {\bf
(ii)} the spins, being attached to conduction electrons, carry
the information along a nonmagnetic sample where they are manipulated 
through spin-dependent interactions; and {\bf (iii)} the information is
read off at the drain (detector). The successful realization of such 
devices demands resolution of some of the paramount fundamental problems 
in current spintronic research:  all-electrical injection, at {\em room temperature}, of 
fully spin-polarized currents into a semiconductor~\cite{semi,injection} (on the same 
footing as demonstrated in metallic devices~\cite{jedema} where large currents of cold-electrons  
can be injected and detected via Ohmic contacts); efficient spin detection at a drain electrode  
(spin measurement is impeded by the same feature that generates long coherence times making it 
technologically interesting---weak coupling to the environment~\cite{spin_measure}); and 
the engineering of spin-orbit couplings.~\cite{nitta}

The paper commences  with the discussion of different types of quantum coherence 
that can be encountered in spintronic structures (Sec.~\ref{sec:hamilton}). These intuitive 
notions are then rigorously elaborated within the tensor product formalism for the joint 
description of spin and orbital quantum states, thereby making it easy to include effects of 
quantum coherence in the computation of transport quantities relevant for spintronics. The 
central such quantity is the conductance matrix ${\bf G}$  describing spin-resolved transport 
measurements. We then introduces an efficient technique to compute ${\bf G}$ by developing 
an implementation (based on lattice Green functions~\cite{datta,nikolic_jcmp} formulated in the 
tensor product of real and spin space, rather than wave functions) of the  Landauer-B\" uttiker 
formulation of linear response transport. Our formalism can accommodates both local and non-local effects of 
spin dynamics. We then reexamine in Sec.~\ref{sec:ucf} the salient quantum interference 
effects of mesoscopics---conductance fluctuations and (anti)localization---for the matrix 
elements of ${\bf G}$, as well as their sums, corresponding to different spin-resolved measurements 
on a ferromagnet-semiconductor-ferromagnet (FM-Sm-FM) system where disorder is gradually 
increased in the central non-magnetic region. This allows us to sweep through 
different transport regimes:  from ballistic (where few scattering events 
take place within the sample of size $L$ since mean free path satisfies 
$\ell \gtrsim L$) to strongly disordered (where mean free path loses its meaning). 
Recent quantum-coherent treatments of spin-polarized transport in two-terminal devices 
have been confined to clean samples~\cite{mireles,matsuyama} because the Datta-Das 
original spin-FET proposal (which has ignited much of the interest for semiconductor 
spintronics) requires strictly ballistic transport regime for the charge degrees of 
freedom. However, some level of scattering (from impurities, boundaries, or at the FM-Sm interface) 
is inevitable, and it is important to understand properties of spin-polarized injection, 
conduction, and degradation of well-defined spin-polarization in such environment.~\cite{disorder_spin_fet} 
For example, peculiar quantum interference effects, arising from the entanglement of spin and 
orbital degrees of freedom,~\cite{seba,hu} are revealed in our disordered FM-Sm-FM system 
when analyzing the relationship between partial conductances (elements of ${\bf G}$) as a 
function of the Rashba SO coupling, disorder strength, and orientation of incoming 
electron spin with respect to the Rashba electric field (Sec.~\ref{sec:conductance}). We summarize 
our main findings in Sec.~\ref{sec:conclusion}.

\section{Hamiltonian approach to quantum spin-polarized transport}\label{sec:hamilton}

\subsection{Quantum coherence within the tensor product formalism}\label{sec:tensor}

In the general formalism of nonrelativistic quantum mechanics,
spin degrees of freedom are internal, and are therefore described
within a separate finite-dimensional vector space ${\mathcal H}_{\rm s}$ which 
has to be multiplied tensorially with an orbital space ${\mathcal H}_{\rm
o}$ to get the {\em full} Hilbert space of quantum states ${\mathcal
H}={\mathcal H}_{\rm o} \otimes {\mathcal H}_{\rm s}$. Any
operator acting in ${\mathcal H}$ can be expressed as a linear
combination of the tensor products $\hat{O}_{\rm o} \otimes
\hat{O}_{\rm s}$, where $\hat{O}_{\rm o}$ and $\hat{O}_{\rm s}$
act in ${\mathcal H}_{\rm o}$ and ${\mathcal H}_{\rm s}$,
respectively. A generic Hamiltonian model for spintronics 
$\hat{H}=\hat{H}_{\rm o} + \hat{H}_{\rm s} + \hat{H}_{\rm so}$ contains 
the following pieces: $\hat{H}_{\rm o} \equiv \hat{O}_{\rm o} \otimes \hat{I}_{\rm s}$ 
which acts nontrivially in ${\mathcal H}_{\rm o}$ and like a unit operator
$\hat{I}_{\rm s}$ in ${\mathcal H}_{\rm s}$; $\hat{H}_{\rm s}
\equiv \hat{I}_{\rm o} \otimes \hat{O}_{\rm s}$ acting
nontrivially in ${\mathcal H}_{\rm s}$; and SO terms
$\hat{H}_{\rm so}$ which are linear combinations of a tensor
product of spin and orbital operators coupling the two
spin-polarized subsystems. The advantage of spintronics modeling 
through a quantum transport formalism that takes as an input a microscopic 
single-particle Hamiltonian is that spin, as a genuine quantum attribute 
of an electron, is treated in the most natural way by an operator acting in a 
two-dimensional vector space ${\mathcal H}_{\rm s}$. The spin-dependent interactions are 
then accounted for by the terms in a Hamiltonian that are functions of the spin operator 
$\hat{\bm S} = \hbar \hat{\bm{\sigma}}/2$ [$\hat{\bm{\sigma}}= (\hat{\sigma}_x,\hat{\sigma}_y, \hat{\sigma}_z)$ 
is a vector of the Pauli spin matrices] acting in ${\mathcal H}_{\rm s}$. An arbitrary state  
vector $|\Psi \rangle \in {\mathcal H}$ is a linear combination of the separable (i.e., uncorrelated) 
states forming a basis $|\phi_\alpha \rangle \otimes |\sigma \rangle$,  where $|\phi_\alpha \rangle$ are 
basis states in ${\mathcal H}_{\rm o}$ and corresponding basis state $|\sigma \rangle$ in ${\mathcal H}_{\rm
s}$ are the spin eigenstates $|\sigma \rangle$ of $\hat{\bm{\sigma}} \cdot \hat{\bf u}$
($\sigma=\uparrow,\downarrow$ is a quantum number labeling the two eigenstates, 
and $\hat{\bf u}$ is a unit vector along the spin-quantization axis which is taken to be parallel to the 
magnetization of the ferromagnetic contacts in FM-Sm-FM geometry).

Most of quantum-interference phenomena observed in standard mesoscopic experiments 
(such as, weak localization, conductance fluctuations, or the Aharonov-Bohm effects~\cite{mesophys}) 
probe the {\em orbital} quantum coherence of electron states, which requires  preservation 
of the relative phase of linear superpositions of spatial states of the electron.~\cite{mesophys} 
Such mesoscopic systems are smaller than the dephasing length $L_\phi$ (typically $L_\phi \lesssim 1$ $\mu$m at 
low enough temperatures $T \lesssim 1K$) set by inelastic processes. Thus, coherence times relevant 
for these investigations are completely disconnected from the spin coherence times $T_2$ over 
which the relative phase in superpositions of $\uparrow$ and $\downarrow$ spin states is well-defined~\cite{qc}  
(transport of phase-coherent spin states has been observed~\cite{kikkawa} over length scale $L_2 \sim 100$ $\mu$m in 
semiconductors at low temperatures where $T_2$ can reach 100 ns). For an ensemble of spins created in experiments~\cite{kikkawa_98}, 
the operational loss of phase coherence is described by the dephasing time $T_2^*$ due to inhomogeneities in the Zeeman terms 
(although such dephasing is independent of individual spin decoherence, it provides experimentally measurable lower bound 
$T_2^* \le T_2$). The other basic time scale which, together with $T_2$, captures essential features of spin dynamics is the
spin relaxation time $T_1$ (typically $T_1 \ge T_2$).  The time $T_1$ is classical in nature since it determines lifetime of 
an excited spin state (aligned long the external magnetic field), i.e., the relaxation of unbalanced populations of spin states 
toward thermal equilibrium so that no concept of superpositions of quantum states is involved.~\cite{jaro_spin} Corresponding spin 
diffusion length $L_1$ sets the upper limit to the size for any spintronic device since beyond this scale spin-encoded information 
completely fades away.~\cite{jaro_spin} The principal interactions providing modes of decoherence and spin relaxation are 
exchange coupling  with other electron (or nuclear) spins and SO coupling to impurity atoms and defects. In spintronic 
systems of  current interest, one can encounter quantum interference effects stemming from spin coherence~\cite{datta90,meijer}, 
both spin and orbital coherence that are independent of each other,~\cite{yuasa} or intertwined spin and orbital coherence 
when SO interaction are strong.~\cite{marcus_so} 

These different quantum-coherent situations  become transparent after being recast in the formalism 
of tensor products of Hilbert spaces and corresponding operators. In conventional {\em macroscopic} solids  
at finite temperatures one has to use a quantum statistical mechanics description of electrons 
in terms of density operators~\cite{ballentine} (i.e., density matrices when a particular representation 
is chosen)
\begin{equation}
\hat{\rho}_{\rm macro} = \hat{\rho}_{\rm o} \otimes
\hat{\rho}_{\rm s}.
\end{equation}
which are determined by thermal equilibrium. The most general quantum state of a 
spin$-\frac{1}{2}$ particle is accounted for by a density operator~\cite{ballentine,entangle}
\begin{equation}\label{eq:spin_mixture}
\hat{\rho}_s=\frac{1}{2} \, \left( \hat{I}_{\rm s}+{\bf p} \cdot \hat{\bm{\sigma}} \right),
\end{equation} 
where $0 \le |{\bf p}| \le 1$. For example, $|{\bf p}| = 1$ describes a pure quantum state which is 
fully {\em polarized} in the direction of vector ${\bf p}$, while $|{\bf p}| = 0$ 
stands for a non-pure state (the so-called  {\em mixture}) that is completely unpolarized. The 
trivial example is injection of both spin states in equal proportion,  $\hat{\rho}_s=1/2\, (|\!\! \uparrow\rangle\langle 
\uparrow \!\!| + |\!\! \downarrow\rangle \langle \downarrow \!\!|)= \hat{I}_s/2$. 
Coupling of  electron spin to the environment of surrounding electronic or nuclear spins 
causes decoherence of spin quantum state, i.e., appearance of mixtures $|{\bf p}|<1$ of 
spin states instead of a single pure state. In semiclassical spintronics, the spin part 
of the full (non-pure) state becomes a pure state $\hat{\rho}_{\rm s}=|\Sigma \rangle \langle \Sigma|$.
Then, after taking only diagonal elements of $\hat{\rho}_{\rm o}$,
a semiclassical description in terms of the Boltzmann distribution
functions for $\uparrow$ and $\downarrow$ electrons is obtained where 
phase-relations between different states are ignored~\cite{datta} (as applicable 
to systems with $L > L_\phi$). Saying that the sample is 
{\em phase-coherent} in standard {\em mesoscopic} physics has a simple 
meaning in the formal language: an electron is described throughout the 
sample by a single wave function $\Phi({\bf r}) = \langle {\bf r}|\Phi \rangle \in {\mathcal H}_{\rm o}$, i.e.,
the orbital factor gets ``purified'' $\hat{\rho}_{\rm o}=|\Phi \rangle \langle \Phi|$ at low enough temperatures 
where dephasing processes are suppressed below $\mu$m scales (while in usual experiments with unpolarized currents 
the spin factor is $\hat{\rho}_s=1/2\,  \hat{I}_s$). In sample with both orbital and 
spin quantum coherence, the full state is pure $\hat{\rho}_{\rm meso}^2=\hat{\rho}_{\rm meso}$, such as 
\begin{equation}\label{eq:pure}
\hat{\rho}_{\rm meso} = |\Phi\rangle  \langle \Phi| \otimes |\Sigma \rangle \langle \Sigma|, 
\end{equation}
The separability of the full state in Eq.~(\ref{eq:pure}) holds only when scattering events are spin-independent (such as 
those generated by lattice imperfections, phonons, and non-magnetic impurities), thereby leaving the spin 
state of a  traversing electron  unchanged. In this case, the orbital and spin factor states evolve 
independently in a coherent fashion within its own factor subspaces  ${\mathcal H}_{\rm s}$ or ${\mathcal H}_{\rm o}$, 
under the unitary operator generated by the corresponding Hamiltonian. However, in the presence of spin-orbit 
interaction, these two quantum-coherences become intertwined (note that if reduced density operator, obtained 
by tracing over the states in the other factor spaces, is pure, than it must be a factor of the total state
operator~\cite{ballentine}). For example, if separable state with  $|\!\! \uparrow \rangle$ spin 
factor is injected into a device with spin-orbit coupling, it will evolve by unitary quantum 
evolution into a linear superposition  $|\Psi_{\uparrow}\rangle$ of the  separable state 
vectors $|\phi_\alpha \rangle \otimes | \sigma \rangle$  of some basis in ${\mathcal H}$. Such 
superpositions are still {\em pure}, albeit non-separable quantum states, i.e.,  the different degrees 
of freedom (orbital and spin) of one and the the same particle have become {\em entangled}.~\cite{entangle} Thus, in 
this case individual spin or orbital states lose their coherence since they are not pure any more 
(although no real ensemble of different quantum states corresponding to such mixtures exists in the sample,~\cite{zeh}  
contrary to the case in the example of mixed spin states given above). This is analogous to the 
general decoherence mechanism that operates by entangling quantum system to its environment.~\cite{zeh} 
It also explains the formal meaning of `spin-orbit quantum interference effects',~\cite{marcus_so} 
since entanglement is entailed by linear superpositions of states in the full Hilbert space  ${\mathcal H}$. 
Moreover, such entangled states display quantum-mechanical non-locality, usually studied in the 
context of entanglement of two particles (e.g., Einstein-Podolsky-Rosen or Bell states~\cite{ballentine}), 
which manifests here as ``wholistic'' spin effects all throughout the system.~\cite{feve}  When both $\uparrow$ 
and $\downarrow$ particles are injected, the state of an electron is a statistical superposition 
$|\Psi_\uparrow\rangle\langle \Psi_\uparrow| + |\Psi_\downarrow\rangle\langle \Psi_\downarrow|$, rather than the 
coherent one $|\Psi_\uparrow\rangle + |\Psi_\downarrow\rangle$. Thus, the quest for spintronic devices 
that would utilize quantum-coherent dynamics of all degrees of freedom requires to master the control 
of spins without destroying the phase coherence of orbital degrees of freedom~\cite{yuasa} (as is the case 
when many-body interactions with surrounding nuclear spins or other electrons~\cite{gefen} are involved), and 
{\em vice versa}.~\cite{marcus_so}

The effects of quantum-coherence on electronic transport are most easily discussed within 
the Landauer-B\" uttiker scattering approach. To study spin-polarized quantum transport, the basis 
of asymptotic scattering states at the Fermi energy $E_F$ in the leads should be comprised of the 
following vectors
\begin{equation}\label{eq:channel}
\langle {\bf r} |p \rangle^\pm_\sigma =
\Phi_p(y) \otimes \exp(\pm ik_px) \otimes | \sigma \rangle.
\end{equation}
In the spinless case, the orbital factors are usually referred to as
conducting `channels'~\cite{carlo_rmt}; here a `spin-polarized channel' 
$p\sigma$ has a real wave number $k_p > 0$, a transverse wave function
$\Phi_p(\bf r)$ defined by  quantization of transverse momentum in the 
leads of finite cross section and given boundary conditions, and a spin factor 
state $|\!\! \uparrow \rangle$ or $|\!\! \downarrow \rangle$. 
Thus, injecting a flux concentrated in the channel $|{\rm in}\rangle \equiv | p \rangle \otimes |\sigma 
\rangle$ will generate the following quantum state in the opposite lead
\begin{equation}\label{eq:t_matrix}
|{\rm out} \rangle = \sum_{q,\sigma^\prime} {\bf
t}_{qp,\sigma^\prime \sigma} |q \rangle \otimes |\sigma^\prime \rangle.
\end{equation}
In the most general situation, with spin-dependent charge scattering present 
in the central region, this state is coherent mixture of all spin-polarized  channels~(\ref{eq:channel}). 
This equation introduces a transmission matrix ${\bf t}$, whose elements  
are ${\bf t}_{qp, \sigma^\prime \sigma}$. The ${\bf t}$-matrix accounts 
for the unitary transformation that the central region would perform on the incoming 
electronic wave function from the left lead.

The formalism  we introduce here is capable of dealing with all of the variety of 
quantum-coherent situations relevant for transport (i.e., coherent superpositions of 
scattering quantum states in ${\mathcal H}_{\rm o}$, ${\mathcal H}_{\rm s}$, or ${\mathcal H}_{\rm o} 
\otimes {\mathcal H}_{\rm s}$) in arbitrary geometry, and for arbitrary values 
of interaction strengths, as long as one stays withing the domain of single-particle 
picture. We choose the Anderson model for $\hat{H}_{\rm o}$ (which is often used for 
numerical modeling of quantum transport,~\cite{datta} localization-delocalization
transition,~\cite{verges} and brute force computation of conductance 
fluctuations,~\cite{ucf_us,nikolic})
\begin{equation}\label{eq:tbh}
  \hat{H}_{\rm o}= \left( \sum_{\bf m} \varepsilon_{\bf m}|{\bf m} \rangle \langle {\bf m}|
  +  \sum_{\langle {\bf m},{\bf n} \rangle}
   t_{\bf mn}|{\bf m} \rangle \langle {\bf n}| \right)
  \otimes \hat{I}_{\rm s},
\end{equation}
on a square $N \times N$ lattice ($L=Na$, with $a$ being the
lattice spacing). Here $t_{\bf mn}=t e^{2\pi i \phi_{\bf mn}}$,
$t$ is the nearest-neighbor hopping integral between $s$-orbitals
$\langle {\bf r}|{\bf m}\rangle = \psi({\bf r}-{\bf m})$ on
adjacent atoms located at sites ${\bf m}=(m_x,m_y)$ of the
lattice. The {\em local orbital basis} is advantageous since it allows
one to treat an {\em arbitrary} measurement geometry, which is an essential 
ingredient in the studies of quantum transport through finite-size mesoscopic 
systems.~\cite{imry} In the applied magnetic field, the link fields 
$\phi_{\bf mn}=-\phi_{\bf nm}$ entering the Peierls phase factor introduce Aharonov-Bohm 
phases into the tight-binding model so that the flux through a given loop $S$ is
$\Phi_S=\sum_{\langle {\bf m},{\bf n} \rangle \in S} \phi_{\bf
mn}$ in units of the flux quantum $h/e$. The disorder is
simulated by taking a random on-site potential with
$\varepsilon_{\bf m}$ being uniformly distributed over the
interval $[-W/2,W/2]$ [the other possibilities to introduce
disorder is random hopping, $t \in [1-2W,1]$, or random fluxes 
$\phi_{\bf mn} \in [0,2\pi)$]. The lattice site states of
$\hat{H}_{\rm o}$ multiplied tensorially by $|\sigma \rangle$
define a basis $|{\bf m}\rangle \otimes |\sigma \rangle$ in
$\mathcal{H}$, and therefore a real$\otimes$spin-space
representation of the operators in our algorithm.

The purely spin part of $\hat{H}$  is exemplified by a Zeeman term
\begin{equation}\label{eq:zeeman}
  \hat{H}_{\rm s}= - \hat{I}_{\rm o} \otimes \mu \hat{\bm{\sigma}}
\cdot {\bf B},
\end{equation}
where $\mu=g^*\mu_B/2$ ($g^*$ is an effective gyromagnetic ratio
and $\mu_B$ is the Bohr magneton) and ${\bf B}$ is the applied
magnetic field. Finally, the relevant SO couplings in 2D
heterostructures~\cite{halperin} (neglecting $\propto v^3$ terms)
are given by the Rashba term~\cite{rashba} $\alpha_{\rm R} m^* (
\hat{\bm{\sigma}} \times \hat{\bf v} )\cdot \hat{\bf z}/\hbar$
($\hat{\bf z}$ is a unit vector orthogonal to the plane of the
sample and  $m^*$ is the effective mass), and the Dresselhaus
term~\cite{halperin} $\eta m^* (\hat{v}_x \hat{\sigma}_x -
\hat{v}_y \hat{\sigma_y})/\hbar$ (generated by the lack of
inversion symmetry of a periodic crystal potential in the bulk,
like in GaAs):
\begin{eqnarray} \label{eq:rashba}
\hat{H}_{\rm so} & = & \frac{ \alpha_{\rm R} \hbar}{2a^2 t}
(\hat{v}_x \otimes \hat{\sigma}_y - \hat{v}_y \otimes
\hat{\sigma}_x) \nonumber \\ && \mbox{} -
\frac{\eta \hbar}{2a^2t}(\hat{v}_x \otimes \hat{\sigma}_x - \hat{v}_y
\otimes \hat{\sigma}_y),
\end{eqnarray}
where we replace $m^*$ by its tight-binding description in terms
of $t=-\hbar^2/2m^*a^2$. The explicit use of the tensor-product 
notation in our formulas makes it possible to obtain quickly and 
efficiently their matrix representation in a chosen basis. For 
example, to obtain the site-spin representation of the Rashba term 
in Eq.~(\ref{eq:rashba}) we use the fact that velocity operator is 
a matrix \begin{equation}
\langle {\bf m} |\hat{v}_x| {\bf n} \rangle = \frac{i}{\hbar}\, t_{\bf  mn} \left( m_x - n_x \right),
\end{equation}
in the site representation, and then multiply, in the sense of direct product 
of matrices (when particular representation is chosen, $\otimes$ stands for the 
Kronecker or direct product of matrices), this matrix with respective 
Pauli matrix. Thus, the final matrix elements in Eq.~(\ref{eq:rashba}) are 
just dimensionless constants multiplied by the material-specific ``spin-orbit 
hopping parameters'' $t_{\rm so}^{\rm R}=\alpha_{\rm R}/2a$ and $t_{\rm so}^{\rm D}=\eta/2a$ 
that set the energy scales of the Rashba and Dresselhaus terms, respectively. 

The Rashba term added to the tight-binding Hamiltonian~(\ref{eq:tbh}) will 
cause spin splitting [i.e., lifting of spin degeneracy for $\uparrow$ and 
$\downarrow$  states having the same momentum, ${\bf k}=(k_x,k_y) \neq 0$] of an energy 
subband of 2DEG into two branches
\begin{eqnarray}\label{eq:spin_split}
E(k_x,k_y) & = &  E_0(k_x,k_y)  \pm \Delta E(k_x,k_y) \nonumber \\
& = & \varepsilon + 2t[\cos (k_xa) + \cos (k_ya)] \nonumber \\ 
&& \pm 2t_{\rm so}^{\rm R}\sqrt{\sin^2(k_xa)+\sin^2(k_ya)}, 
\end{eqnarray} 
for clean sample with constant $\varepsilon$ on the diagonal 
[the splitting becomes linear~\cite{mireles} in the momentum $\Delta E=2\alpha_{\rm R}  
\sqrt{k_x^2 + k_y^2}$  for parabolic subband dispersion $E_0(k_x,k_y)=\hbar (k_x^2 + k_y^2)/2m^*$ 
characterizing continuous systems]. Such splitting of the conduction band as a result 
of spin-orbit coupling in the presence of an asymmetric confinement potential makes it a useful 
tool to model the electronic structure of confined narrow-gap semiconductors. Rashba SO term also 
induces, when viewed within the semiclassical picture, spin precession around effective Rashba magnetic 
field ${\bf B}_R({\bf k})$  [Rashba Hamiltonian, like any SO term,  can be interpreted as the interaction 
of electron spin with  ${\bf k}$-dependent internal effective magnetic field, ${\bf B}({\bf k})\cdot \hat{\bf S}$] 
with frequency $\omega = \Delta E/2\hbar$. In the presence of disorder, the precession around ${\bf B}({\bf k})$ 
for a given ${\bf k}$ terminates after scattering of impurity (the other sources of scattering that 
change ${\bf k}$ are boundary surface and phonons). Then is starts again but along a different 
randomly selected axis, so that the change of spin direction by full precession, which 
occurs  in ballistic systems,~\cite{datta90,mireles} is suppressed by disorder. In the cases of our study 
where  interplay of the Rashba interaction and charge scattering takes place (at zero temperature), initial 
full polarization of injected electrons is destroyed (technically, measurement of the spin-dependent properties 
alone requires to use reduced density operator, which does not describe a pure state  any more). This is 
analogous to  D'yakonov-Perel' mechanism,~\cite{dyakonov} which is usually discussed in the context of spin 
relaxation~\cite{jaro_spin}, but in our study of transport through mesoscopic spintronics structures, which are 
effectively at zero-temperature, it is responsible for the spin decoherence.~\cite{entangle}

\subsection{Landauer-type conductance formula for quantum spin-polarized transport}\label{sec:landauer}

The central result of our formalism is a direct algorithm to compute exactly the zero-temperature 
conductance matrix~\cite{seba} of a single sample 
\begin{equation} \label{eq:qc}
    {\bf G} =
     \left( \begin{array}{cc}
     G^{\uparrow\uparrow} &  G^{\uparrow\downarrow} \\
      G^{\downarrow\uparrow} &  G^{\downarrow\downarrow}
  \end{array} \right) =\frac{e^2}{h} {\bf T}=\frac{e^2}{h}
      \left( \begin{array}{cc}
     T^{\uparrow\uparrow} &  T^{\uparrow\downarrow} \\
      T^{\downarrow\uparrow} &  T^{\downarrow\downarrow}
  \end{array} \right),
\end{equation}
We imagine that the 2D sample is attached via perfect Ohmic contacts 
(which are desirable but are currently problematic to achieve at FM-Sm 
interface together with efficient injection~\cite{injection}) to two ideal 
semi-infinite leads (Fig.~\ref{fig:setup}) where one serves to inject 
spin-polarized current from ferromagnetic contacts (emitting
$\uparrow$ or $\downarrow$ electrons) into a semiconductor or
nonmagnetic metal,~\cite{johnson}  and the other one drains
electrons to the contacts detecting $\uparrow$ and/or $\downarrow$
electrons. The leads also define, by transverse
quantization, the orbital part of an asymptotic scattering state in Eq.~(\ref{eq:channel}).
Moreover, they effectively remove ``hot'' electrons from the open
sample, thereby bypassing the intricate issues of dissipation
which must occur somewhere in the circuit to reach the
steady-state regime under external pumping. While the trick of
using ideal semi-infinite leads is a standard one in mesoscopic
transport theory,~\cite{datta}  the polarization of the
asymptotic states stems from the different densities of states
for $\uparrow$ and $\downarrow$ electrons in the ferromagnetic contacts 
(which can be conventional metallic or semiconducting, currently 
available only at low temperature~\cite{ohno_dsm}). The meaning of 
${\bf G}$ is elucidated by casting it into a form analogous to the  
Landauer formula,~\cite{datta,carlo_rmt} where the usual sum of the
transmission probabilities over all transverse propagating modes
for spinless particles  $T = {\rm Tr}\, {\bf tt}^\dag$ is
replaced by a $2 \times 2$ matrix  ${\bf T}$ of partial
transmission coefficients describing the transition  between left
and right subsystems comprised of the two types $\uparrow$,
$\downarrow$ of spin-polarized asymptotic states.~\cite{seba}
Thus, $T^{\uparrow\uparrow}$ is the sum of transmission
probabilities, over all conducting channels, for $\uparrow$ 
electron to travel from the left to the right lead whose magnetic
moments are oriented  in parallel. Analogously, $T^{\uparrow\downarrow}$ 
quantifies the probability to detect (in the setup with antiparallel 
magnetic moments of the contacts) $\downarrow$-electron arising from 
spin precession~\cite{datta90,mireles}  or spin-flips~\cite{seba} of 
the injected $\uparrow$-electrons.

Diagonal components of ${\bf G}$, $G^{\uparrow\uparrow}$ and
$G^{\downarrow\downarrow}$, are familiar from the studies of 
giant-magnetoresistance~\cite{gijs} (in the case of spin-degenerate 
transport, trivially, $G^{\uparrow \uparrow}=G^{\downarrow \downarrow}$ 
and $G^{\uparrow \downarrow}=G^{\downarrow \uparrow}=0$). On the other 
hand, properties of the off-diagonal ``mixing'' conductances,
$G^{\uparrow \downarrow}$ and $G^{\downarrow \uparrow}$, are much
less explored (also, for some problems of partially polarized
transport between non-collinear ferromagnets they have been
introduced recently as complex quantities~\cite{brataas}). The
${\bf G}$-matrix allows one to compute the measured conductance
in different setups for spin-resolved experiments
\begin{equation}\label{eq:measured_g}
G_{\rm measured}={\bf s}_{\rm i}^\dag \cdot {\bf G} \cdot {\bf s}_{\rm c},
\end{equation}
where ${\bf s}_{\rm i},\, {\bf s}_{\rm c}$ are vectors describing
the type of incoherent mixture of $\uparrow$ or $\downarrow$ electrons which
are injected or collected, respectively. This quantum conductance
expression can also be employed to get results of semiclassical spintronics
by taking a disorder average and the semiclassical approximation in the large
system limit, as has been the practice in related
giant-magnetoresistance studies of the transport with 
two independent `spin-channels'.~\cite{gijs}  

The conductance matrix Eq.~(\ref{eq:qc}) is calculated by
generalizing a two-probe Landauer-type formula for the
spin-degenerate transport, which is obtained in the
linear-response and zero-temperature limit of an expression
derived by the Keldysh technique,~\cite{caroli,meir} to a
separate treatment of the two spin-polarized components
\begin{subequations}
\label{eq:landauerspin}
\begin{eqnarray}
  {\bf G} & = & \frac{e^2}{h} \, \sum_{i,j=1}^N
     \left( \begin{array}{cc}
      |{\bf t}_{ij,\uparrow\uparrow}|^2 &  |{\bf t}_{ij,\uparrow\downarrow}|^2 \\
      |{\bf t}_{ij,\downarrow\uparrow}|^2 & |{\bf t}_{ij,\downarrow\downarrow}|^2
  \end{array} \right), \\
  {\bf t} & = & 2 \sqrt{-\text{Im} \, \hat{\Sigma}_L\otimes
\hat{I}_{\rm s} } \cdot \hat{G}^{r}_{1 N} \cdot
  \sqrt{-\text{Im}\, \hat{\Sigma}_R \otimes \hat{I}_{\rm s}}.
\end{eqnarray}
\end{subequations}
The partial summation adds squared amplitudes of all elements of
the transmission matrix ${\bf t}$ having the same spin indices
\begin{subequations}
\begin{eqnarray}
\label{eq:tindices}
    {\bf t}_{ij,\uparrow\uparrow} & \equiv & {\bf t}_{2(i-1)+1,2(j-1)+1}, \\
    {\bf t}_{ij,\uparrow\downarrow} & \equiv & {\bf t}_{2(i-1)+1,2j}, \\
    {\bf t}_{ij,\downarrow\uparrow} & \equiv & {\bf t}_{2i,2(j-1)+1}, \\
    {\bf t}_{ij,\downarrow\downarrow} & \equiv & {\bf t}_{2i,2j},
\end{eqnarray}
\end{subequations}
where $i,j=1,...,N$. Here $-\text{Im} \,
\hat{\Sigma}_{L,R}=-(\hat{\Sigma}_{L,R}^r-
\hat{\Sigma}_{L,R}^a)/2i$ are nonnegative matrices with a
well-defined matrix square root, where
$\hat{\Sigma}_{L,R}^{a}=[\hat{\Sigma}_{L,R}^{r}]^{\dagger}$ are
the self-energies ($r$-retarded, $a$-advanced) describing the
``interaction'' of a sample with the left ($L$) or right ($R$)
lead. The $2N \times 2N$ submatrix $\hat{G}^{r}_{1 N}$ of the full
Green function matrix $\hat{G}^{r}_{{\bf
nm},{\sigma\sigma^\prime}}= \langle {\bf n},\sigma | \hat{G}^{r}
| {\bf m}, \sigma^\prime \rangle$ connects the layers (i.e., rows of sites) 
$1$ and $N$ along the direction of transport ($x$-axis). The Green function,
describing the propagation of $\uparrow$ or $\downarrow$ electron
between two arbitrary sites inside an open conductor in the
absence of inelastic processes, is the site-spin representation
of the Green operator obtained by inverting the Hamiltonian for the 
given boundary conditions (hard wall in our case)
\begin{equation} \label{eq:green}
\hat{G}^{r}=[E \hat{I}_{\rm o} \otimes \hat{I}_{\rm s}-\hat{H}
- \hat{\Sigma}^{r} \otimes \hat{I}_{\rm s}]^{-1},
\end{equation}
where $\hat{\Sigma}^{r}=\hat{\Sigma}_L^{r}+\hat{\Sigma}_R^{r}$.
The self-energy matrices introduced by the leads are non-zero
only on the end layers of the sample adjacent to the leads. Their 
analytical form is known exactly as $\hat{\Sigma}^{r}_{L,R}({\bf
n},{\bf m}) = t_{\rm C}^2 \hat{g}_{L,R}^r({\bf n}_S,{\bf
m}_S,t_{\rm L})$, with $\hat{g}_{L,R}^r({\bf n}_S,{\bf m}_S,t_{\rm
L})$ the surface Green function of the bare semi-infinite
lead between the sites ${\bf n}_S$ and ${\bf m}_S$ in the end
atomic layer of the lead (adjacent to the corresponding sites
${\bf n}$ and ${\bf m}$ inside the
conductor).~\cite{datta,nikolic_jcmp} Here $t_{\rm L}$ and
$t_{\rm C}$ are the hopping parameters in the lead  and on the
lead-sample interface, respectively (see Fig.~\ref{fig:setup}).

Although Eq.~(\ref{eq:landauerspin}) assumes dynamics of
noninteracting quasiparticles, it is also valid for the 
zero-temperature linear-response regime of an interacting
system~\cite{meir} (only single-electron elastic processes are
allowed in this regime due to the requirements of energy
conservation). In the scattering picture of transport of
noninteracting quasiparticles on the tight-binding lattice, 
the orbital factor of the asymptotic states Eq.~(\ref{eq:channel}) is
comprised of a quantized transverse wave function and a Bloch state
(instead of a plane wave), where the corresponding dispersion relation
is defined by the semi-infinite tight-binding lattice modeling the
leads.~\cite{nikolic_jcmp} Since the Rashba electric field selects a
preferred direction in space, one has to take into account the
relative spin orientation of the incoming electron with respect
to the $z$-axis. We choose the representation for the spin-part of the 
scattering state in Eq.~(\ref{eq:channel}) as $|\!\! \uparrow
\rangle \rightarrow \left( \begin{array}{c}
       1  \\
       0
  \end{array} \right)$ and $|\!\! \downarrow \rangle \rightarrow \left( \begin{array}{c}
       0  \\
       1
  \end{array} \right)$. Therefore, the direction of the spin-polarization of 
injected and collected electrons is defined by specifying the direction for 
which spin-operator is diagonal. For example, to study the injection of 
particles that are spin-polarized along $x$-, $y$- and $z$- axis, we change 
the representation of the Pauli spin operators. For $\uparrow$, $\downarrow$ along 
the $z$-axis
\begin{equation}
 \hat{\bm{\sigma}}(Z)  =   \left[   \left( \begin{array}{cc}
       0 & 1 \\
       1 & 0
  \end{array} \right),
   \left( \begin{array}{cc}
       0 & -i \\
       i & 0
  \end{array} \right),
    \left( \begin{array}{cc}
       1 & 0 \\
       0 & -1
  \end{array} \right) \right],
\end{equation}
the usual textbook representation, where $\hat{\sigma}_z$ is diagonal, is used; for 
spin-polarization in the $y$-direction
\begin{equation}
 \hat{\bm{\sigma}}(Y)  =   \left[   \left( \begin{array}{cc}
       0 & -i \\
       i & 0
  \end{array} \right),
   \left( \begin{array}{cc}
       1 & 0 \\
       0 & -1
  \end{array} \right),
    \left( \begin{array}{cc}
       0 & 1 \\
       1 & 0
  \end{array} \right) \right],
\end{equation}
and for spin-polarization along the $x$-axis
\begin{equation}
 \hat{\bm{\sigma}}(X)  =   \left[   \left( \begin{array}{cc}
       1 & 0 \\
       0 & -1
  \end{array} \right),
   \left( \begin{array}{cc}
       0 & -i \\
       i & 0
  \end{array} \right),
    \left( \begin{array}{cc}
       0 & -1 \\
       -1 & 0
  \end{array} \right) \right].
\end{equation}
Note that only the diagonal matrix is unique, while the other two matrices 
are affected (as demonstrated here via one particular choice)  by the freedom in 
choosing phase factors of eigenvectors forming the columns of unitary matrices 
$\hat{U}$ which transform  the standard set of Pauli matrices into a new diagonal 
representation $\hat{U}^\dag \hat{\bm{\sigma}} \hat{U}$ (therefore, it is more appropriate to 
talk about Pauli spin algebra whose operators satisfy abstract set of rules~\cite{ballentine}). 
The injection of electrons which are spin-polarized in the direction of transport ($x$-axis) 
and precess within the central region is a standard set-up of the spin-FET proposal. On the other hand, 
injecting $y$-axis (Fig~\ref{fig:setup}) polarized electrons, whose spin is conserved throughout 
the semiconductor, corresponds to spin valves.~\cite{matsuyama}

\section{Conductance fluctuations and antilocalization of spin-polarized electrons in
disordered Rashba spin-split electron system} \label{sec:ucf}

A major boost for the development of mesoscopic physics came from
the pioneering experimental~\cite{webb_exp} and
theoretical~\cite{ucf_us,ucf_rus} studies of unexpectedly large
conductance fluctuations (CF), ${\rm Var}\, G  \sim (e^2/h)^2$ .
Each phase-coherent sample is characterized by a fingerprint of
reproducible (time-independent) conductance fluctuations as a
function of magnetic field, Fermi energy, or change of impurity
configuration at a fixed $E_F$. This violates the traditional
notion that a given sample is well described by average values of
physical quantities, particularly when ${\rm Var}\, G$ becomes of
the same order as $\langle G \rangle$ [$\langle ... \rangle$
denotes disorder averaging] as $L$ approaches the localization
length $\xi$. Thus, even in metals, CF scale as ${\rm Var}\,
G/\langle G \rangle^2 \sim L^{4-2d}$ in $d$-dimensions. Such
counterintuitive behavior of CF and lack of self-averaging
(especially in low-dimensional systems $d \le 2$) necessitated
the development of a ``mesoscopic approach'' as an alternative to 
standard statistical mechanics description of macroscopic condensed matter 
systems, where whole distribution functions of physical quantities are 
to be studied in nanoscale quantum-coherent solids containing  a
\begin{figure}
\centerline{ \psfig{file=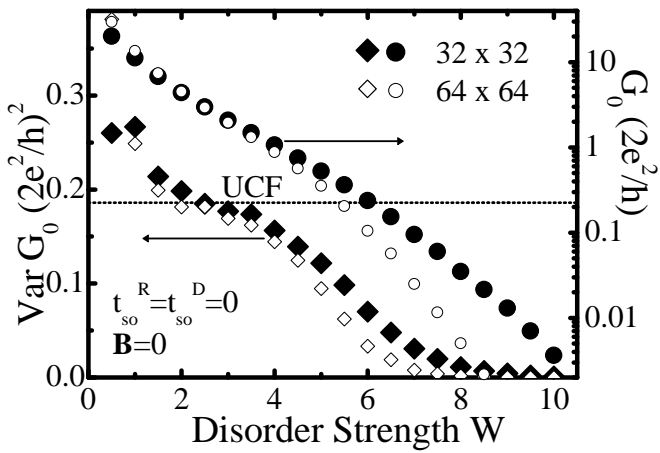,height=3.4in,angle=-90}
} \caption{Conductance $G_0$ (circles) and its variance ${\rm
Var}\, G_0$ (diamonds) in a time-reversal (${\bf B}$=0) and
spin-rotation invariant ($t_{\rm so}^{\rm R}=t_{\rm so}^{\rm
D}=0$) 2D conductor modeled on $32 \times 32$ or $64 \times 64$
half-filled ($E_F=0$) tight-binding lattice as a function of
disorder strength $W$ (disorder-averaging is performed over 10000
samples). The dotted line is the value of the universal
conductance fluctuations predicted
perturbatively~\cite{ucf_us,ucf_rus} for the metallic diffusive
regime.} \label{fig:ucf_spinless}
\end{figure}
macroscopically large number of particles.~\cite{mesophys} In 
diffusive ($\ell \ll L \ll \xi$) metallic ($G \gg e^2/h$)
conductors, CF are considered to be universal (UCF), ${\rm Var}\,
G = C_d (2e^2/h)^2$, where $C_d$ is a constant independent of
the sample size or the degree of disorder,~\cite{carlo_rmt} at
least within certain limits ($C_d$ is reduced by a factor of 2, 4, or
8 by breaking time-reversal or spin-rotation invariance, or
both;~\cite{carlo_rmt} for an exhaustive list of symmetry classes
and possible crossovers in weakly disordered quantum dots and for
weak SO coupling, see Ref.~\onlinecite{aleiner}). However,
numerical studies~\cite{nikolic} of the evolution of ${\rm Var}\,
G$ in 3D mesoscopic metals show that CF monotonically decay with
increasing disorder strength, becoming close to a constant
predicted by the perturbative UCF theory~\cite{ucf_us,ucf_rus}
only in a narrow range of $W$. Therefore, to have a reference
point for the subsequent study of fluctuations of ${\bf G}$, we
first calculate ${\rm Var}\, G_0$ in the crossover between small
and large $W$ (i.e., from a quasiballistic, to a diffusive, and 
finally localized transport regime) for a conductor described by the
non-interacting and spin-independent Hamiltonian $\hat{H}_{\rm
o}$ with ${\bf B}=0=\phi_{\bf mn}$. The localization length in
the band center $E_F=0$ of the paradigmatic Anderson
model~(\ref{eq:tbh}) is~\cite{verges} $\xi \simeq (1+5.2\cdot
10^4/W^4)a$, while its mean free path at the same Fermi energy is
given by $\ell \simeq 30a(t/W)^2$. In this case, spin-degeneracy
gives $G_0=G^{\uparrow \uparrow}+G^{\downarrow \downarrow}=2
G^{\uparrow \uparrow}$ ($G^{\uparrow \downarrow}=G^{\downarrow
\uparrow}=0$). The result plotted in Fig.~\ref{fig:ucf_spinless}
demonstrates that ${\rm Var}\, G_0$ decreases systematically with
increasing $W \in [0,10]$, while $C_2 = 0.186$ is the expected
UCF value in 2D ``metals'' in the diffusive regime~\cite{ucf_us}
(in the quasiballistic regime CF are not expected to be
universal~\cite{asano}).

A recent experimental investigation~\cite{marcus} of transport
through open ballistic (but chaotic due to the surface scattering)
quantum dots in a GaAs heterostructure, that are exposed to a
large in-plane magnetic field, point to the emergence of a 
Rashba SO interaction as the essential ingredient in analyzing
their CF.~\cite{halperin,aleiner} Inspired by their unexpected
findings on the reduction of CF, we undertake
\begin{figure}
\centerline{ \psfig{file=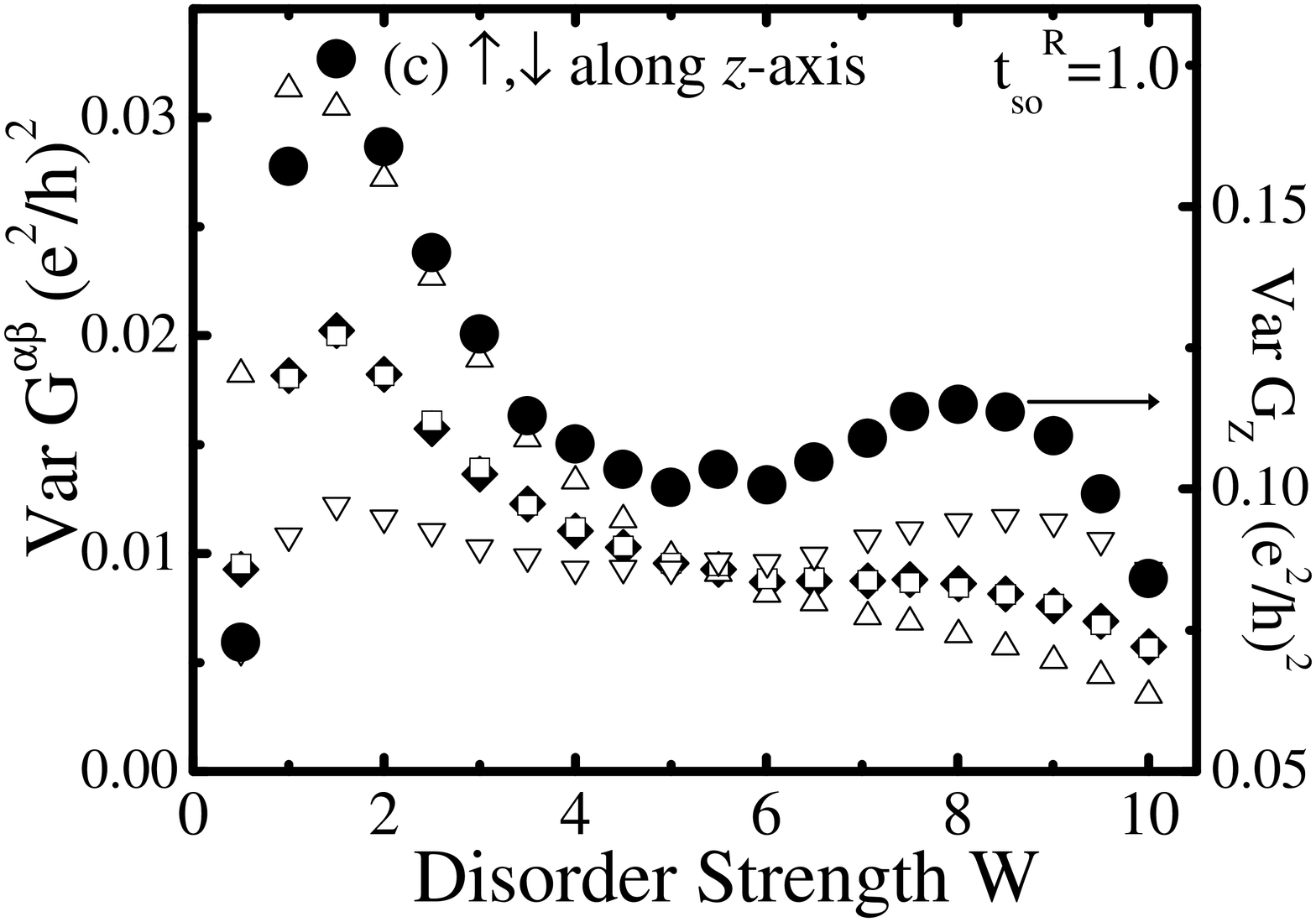,height=3.4in,angle=-90} }
\vspace{0.1in} \centerline{
\psfig{file=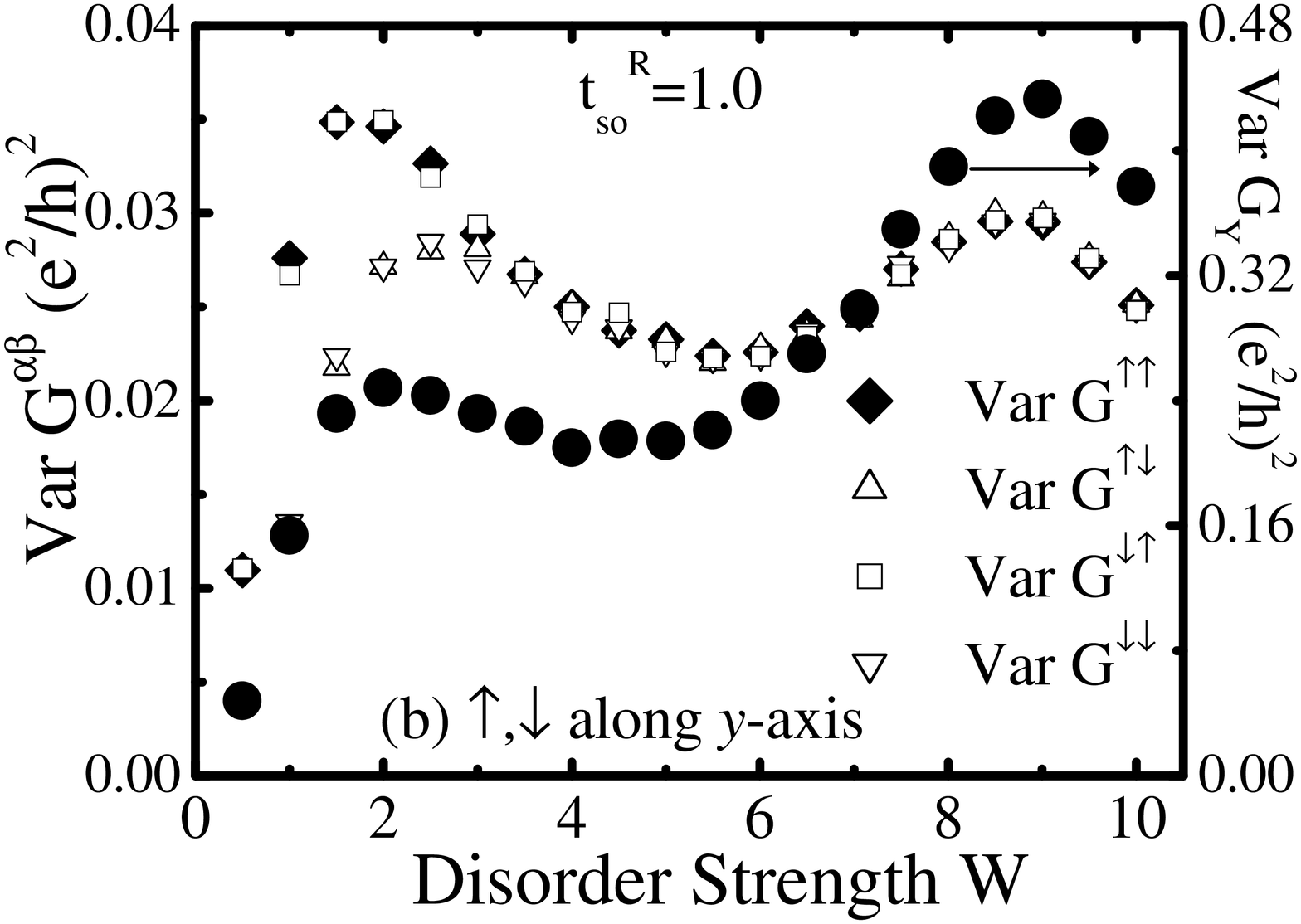,height=3.4in,angle=-90} }
\vspace{0.1in} \centerline{
\psfig{file=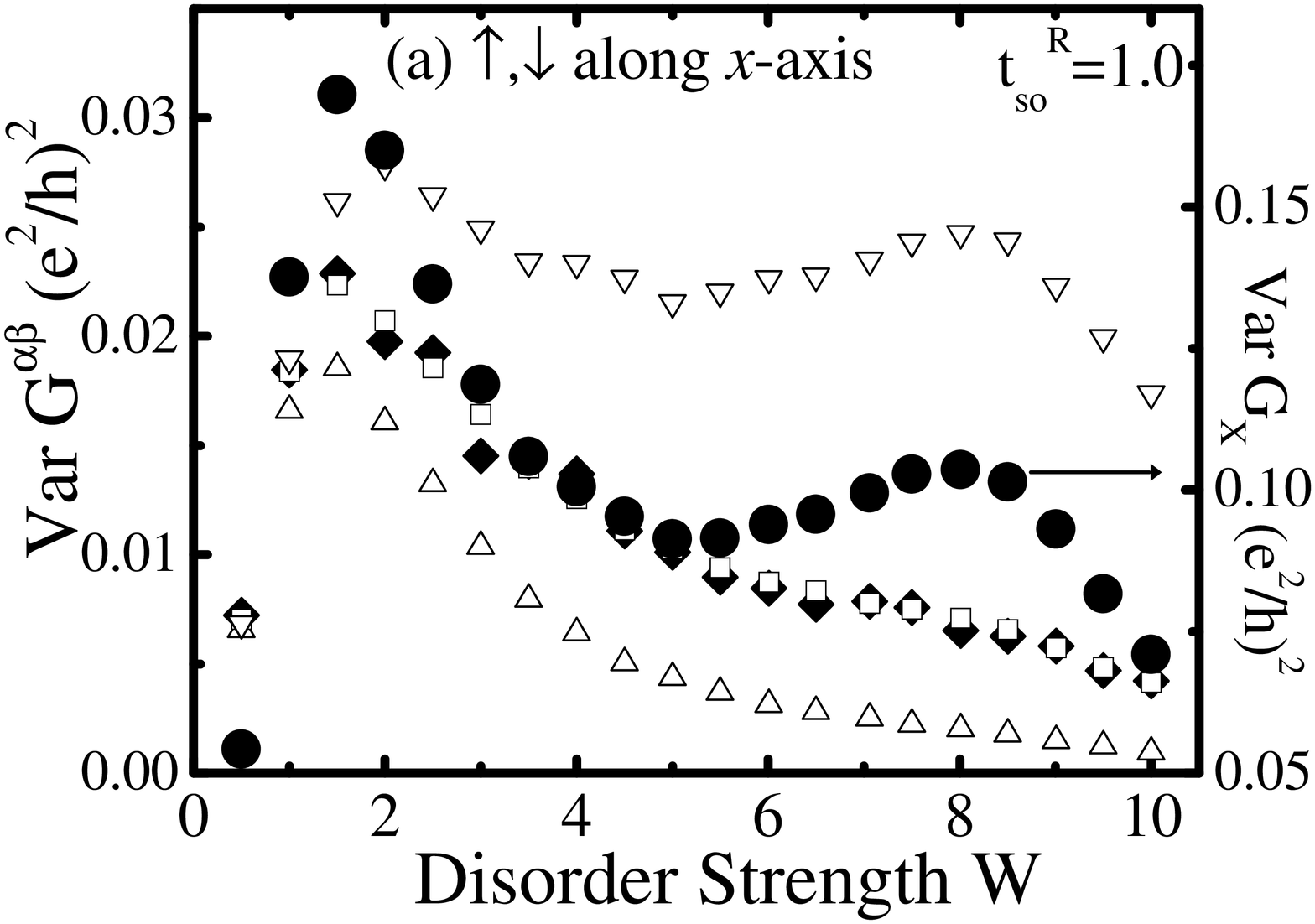,height=3.4in,angle=-90} }
\caption{Zero-temperature sample-to-sample fluctuations of the
components of ${\bf G}$ for transport which is spin-polarized
$\uparrow, \, \downarrow$ along: (a) the $x$-axis, (b) the
$y$-axis, and (c) the $z$-axis. The 10000 samples are modeled on a
$32 \times 32$ half-filled ($E_F=0$) tight-binding lattice with
diagonal disorder and a strong $t_{\rm so}^{\rm R}=1.0$ Rashba SO
coupling. The circles on all three panels are conductance
fluctuations of the sum of the components of ${\bf G}$, e.g.,
${\rm Var} \, G_Z ={\rm Var}\, [ G^{\uparrow\uparrow} +
G^{\downarrow\downarrow} + G^{\uparrow\downarrow} +
G^{\downarrow\uparrow}]$, which describes injection and detection
of both spin species polarized in the $x$-, $y$- or $z$-direction
for ${\rm Var}\, G_X$, ${\rm Var}\, G_Y$, and  ${\rm Var}\, G_Z$,
respectively.} \label{fig:ucf_spin}
\end{figure}
\begin{figure}
\centerline{
\psfig{file=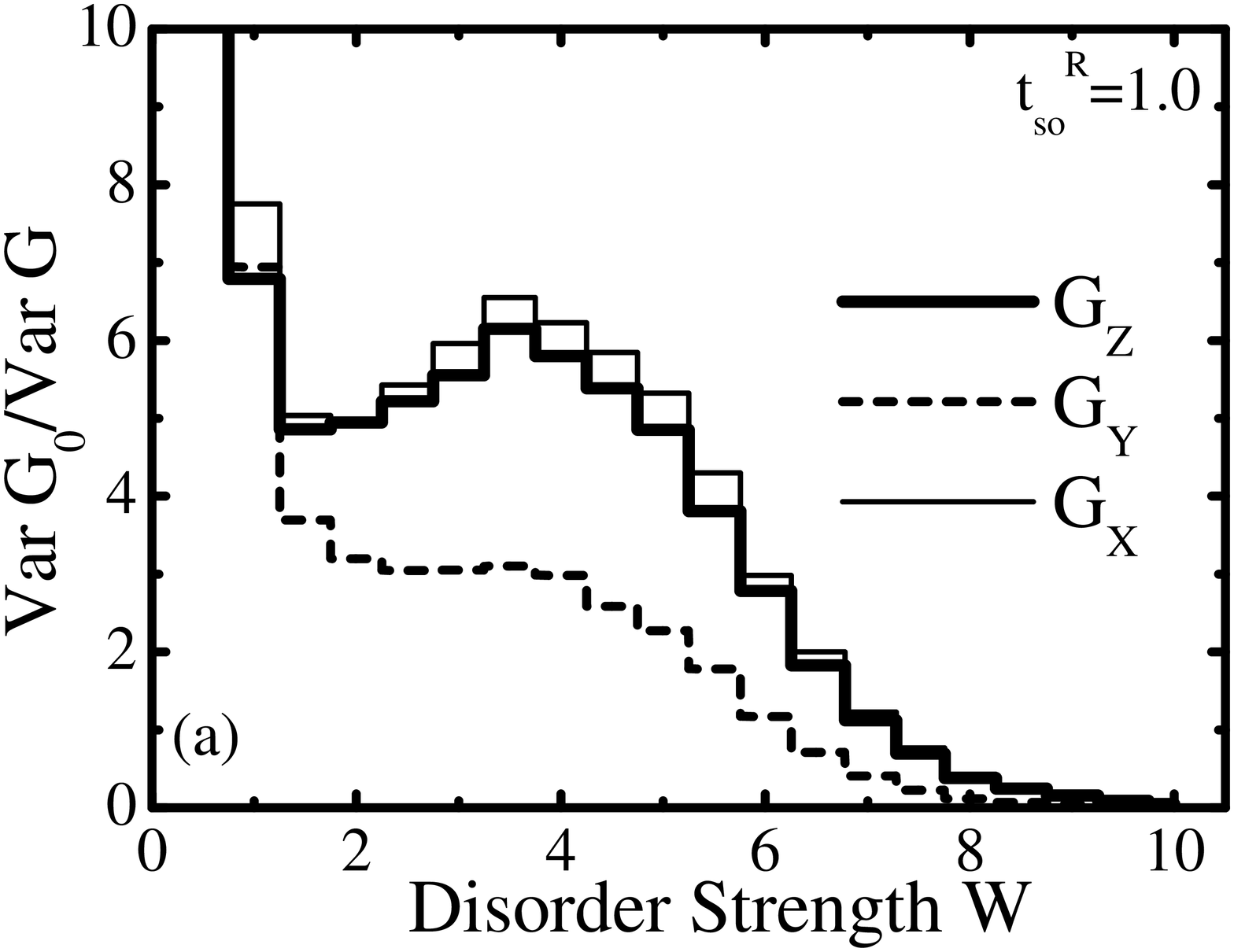,height=3.2in,angle=-90} }
\vspace{0.1in} \centerline{
\psfig{file=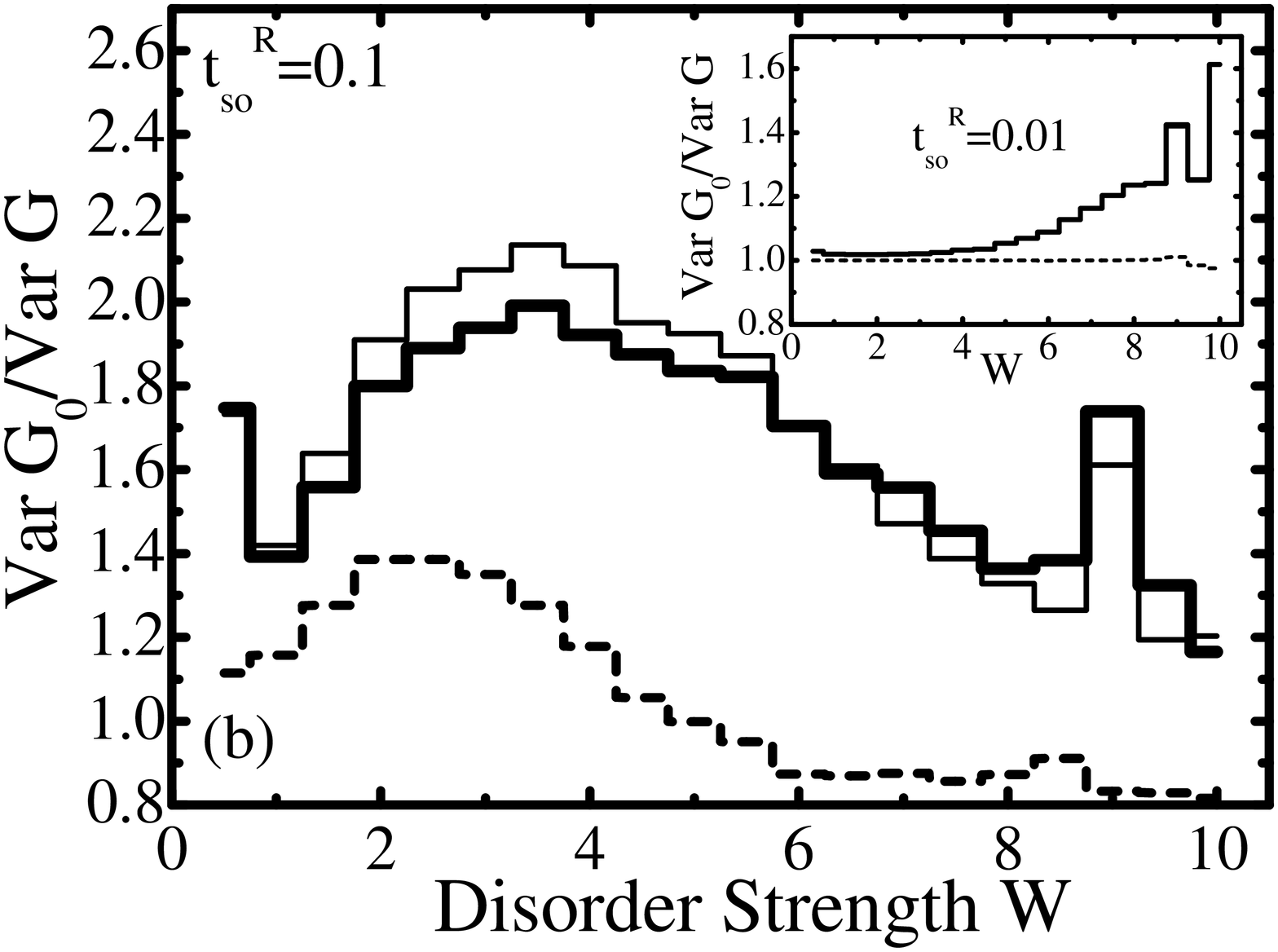,height=3.2in,angle=-90} }
\caption{Ratio ${\rm Var}\, G_0/{\rm Var}\, G$ of the conductance
fluctuations in a pure disordered case
(Fig.~\ref{fig:ucf_spinless}) to the fluctuations of the sums of
the elements of ${\bf G}$  ($G \equiv G_X$, $G_Y$, and $G_Z$)
from Fig.~\ref{fig:ucf_spin}. This reduction factor due to the Rashba
interaction is plotted for: (a) the strong SO coupling limit $t_{\rm
so}^{\rm R}=1.0$; and (b) the weak SO coupling limit $t_{\rm so}^{\rm
R}=0.1$ or 0.01.} \label{fig:reduction}
\end{figure}
a similar but spin-resolved analysis when spin-polarized electrons
are injected into a 2D disordered open quantum dot. Thus, we
``create'' an experiment on the computer ``measuring'' the
components of ${\bf G}$ for each realization of disorder, and
thereby sample-to-sample fluctuations ${\rm
Var}\,G_{\alpha \beta}=\langle G_{\alpha\beta}^2 \rangle - \langle
G_{\alpha\beta} \rangle^2$ as a function of the disorder strength
$W$ and Rashba spin-orbit hopping parameter $t_{\rm so}^{\rm R}$.
Since the Rashba SO coupling can be tuned in principle by an
interface electric field,~\cite{nitta} we sweep $t_{\rm so}^{\rm
R}$ (and neglect $t_{\rm so}^{\rm D}$) from the week $t_{\rm so}^{\rm
R}=0.01$ to the strong $t_{\rm so}^{\rm R}=1.0$  SO interaction
limit~\cite{rashba_values} (a similar range of Rashba hopping has
been explored for spin-polarized transport through clean wires
where it was shown that even in the ballistic case, subbands in 
Eq.~\ref{eq:spin_split} defined by the orbital factor states can be 
mixed depending on the  strength of the Rashba coupling and  wire 
width~\cite{mireles}). The important parameters to keep in mind are: $t_{\rm so}^{\rm R}/W$; 
the ratio $\ell/L$ which delineates boundaries of different transport
regimes in the crossover from ballistic $\ell/L \gg 1$ to
diffusive $\ell/L \ll 1$ transport (e.g., the diffusive transport 
regime can be reached at much weaker disorder but in very large
sample, while the interplay of coherent scattering off impurities
and spin-dependent interactions is the pronounced when the
energy scale of the terms in $\hat{H}$ introduced by
$\hat{H}_{\rm s}$ and $\hat{H}_{\rm so}$ are comparable to the
strength of the disorder); and $\xi/L$ since for $\xi/L < 1$
electrons become localized.
\begin{figure}
\centerline{ \psfig{file=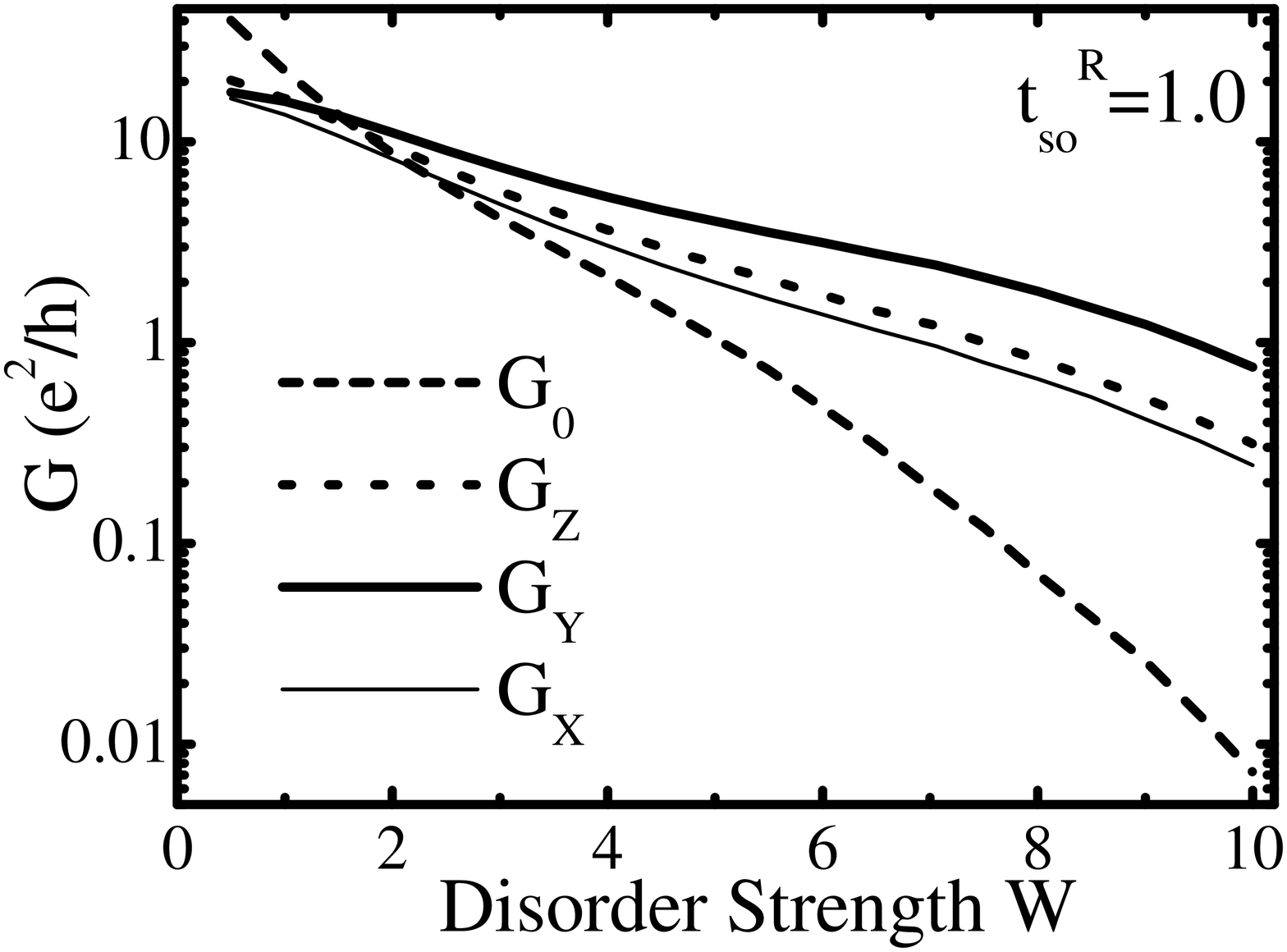,height=3.2in,angle=-90}}
\vspace{0.1in}
\centerline{\psfig{file=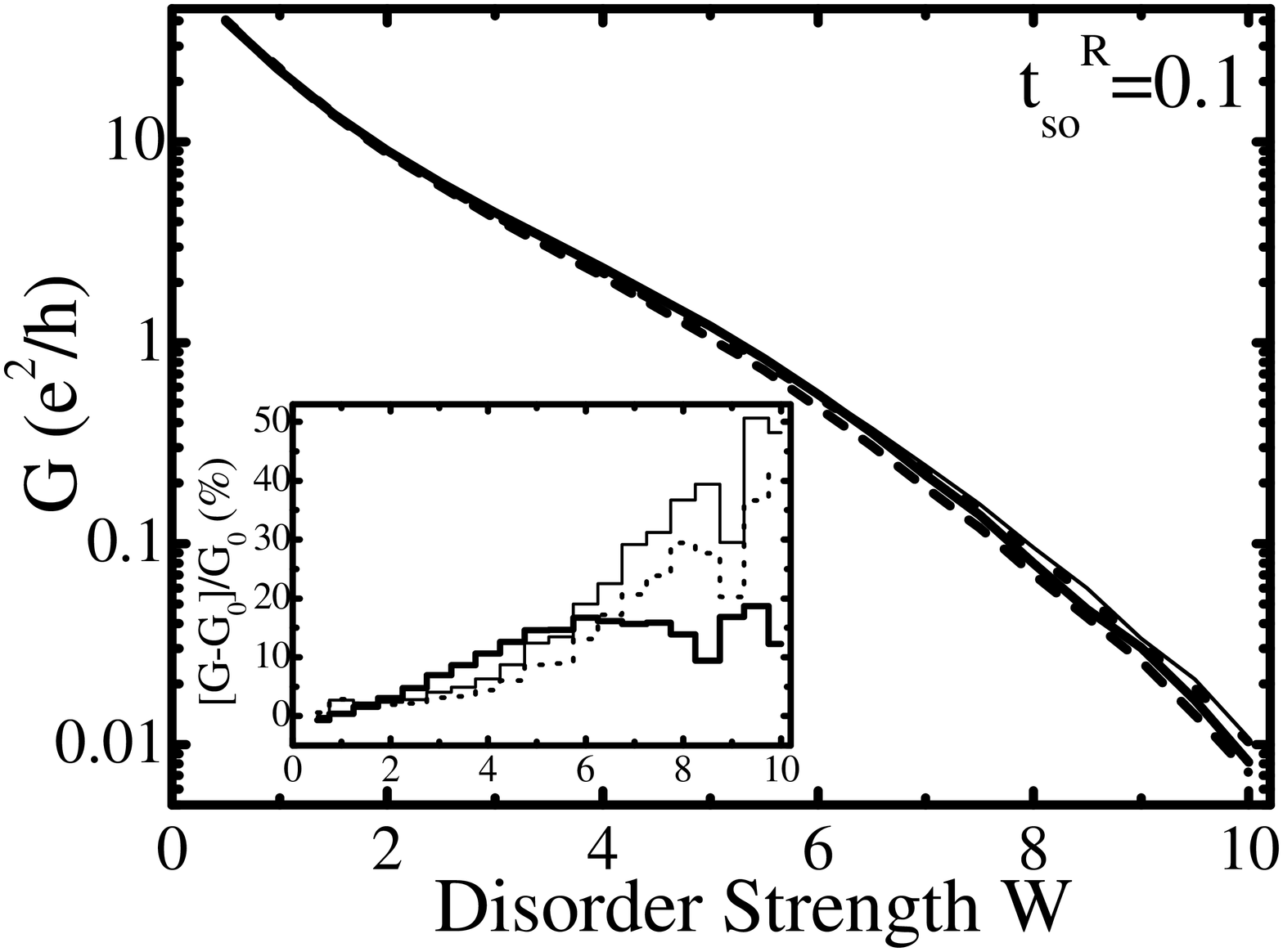,height=3.2in,angle=-90}}
\vspace{0.1in}
\centerline{\psfig{file=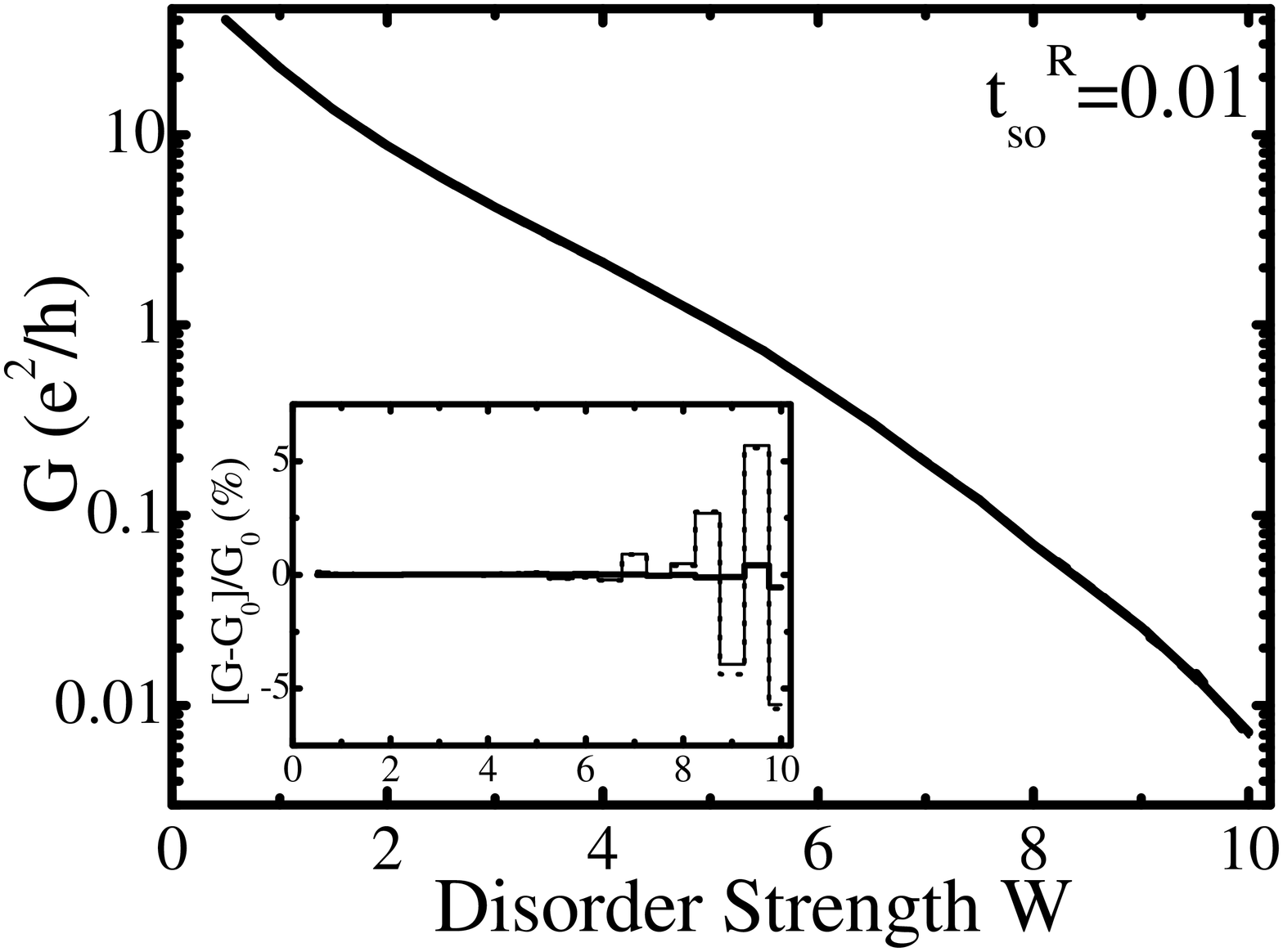,height=3.2in,angle=-90}}
\caption{The total conductance $G_{X,Y,Z} = G^{\uparrow\uparrow}
+ G^{\downarrow\downarrow} + G^{\uparrow\downarrow} +
G^{\downarrow\uparrow}$ describing injection and collection of
both spin-species in the device of Fig.~\ref{fig:setup}.
Disorder-averaging is performed over the same ensemble of samples
studied in Fig.~\ref{fig:ucf_spin}. Antilocalization effects
$G_X$, $G_Y$, $G_Z > G_0$ are substantial (destroying Anderson
insulator formed for $W \protect \gtrsim 6$) at strong Rashba SO
coupling $t_{\rm so}^{\rm R}=1.0$, much smaller for $t_{\rm
so}^{\rm R}=0.1$, and virtually gone at $t_{\rm so}^{\rm
R}=0.01$.} \label{fig:antilocal}
\end{figure}

The mesoscopic fluctuating properties of the components of ${\bf
G}$, for FM magnetization lying $\uparrow$, $\downarrow$ in
the $x$-, $y$- and $z$-direction,  are shown in
Fig.~\ref{fig:ucf_spin} for the largest $t_{\rm so}^{\rm R}=1.0$.
The complete suppression of CF in the weakest disorder case
(standard conductance fluctuations reach a maximum in the
quasiballistic transport regime,~\cite{asano} as shown in
Fig.~\ref{fig:ucf_spinless}) is a feature of the strong SO coupling
limit. While ${\rm Var}\, G^{\uparrow\uparrow} \approx {\rm Var}\,
G^{\downarrow \downarrow}$ in all cases, fluctuations of the
off-diagonal partial conductances ${\rm Var}\,
G^{\uparrow\downarrow}$ and ${\rm Var}\, G^{\downarrow \uparrow}$
can display quite different patterns and even be independent of
the disorder strength (``universality''), unlike the case of
$G_0$ or the total conductances $G_X$, $G_Y$, $G_Z$ introduced
below.

The counterpart of the measured conductance in experiments on
unpolarized electrons~\cite{marcus} would be to inject and collect
both pure spin states ${\bf s}_{\rm i}= {\bf s}_{\rm c}=\left(
\begin{array}{c}
       1  \\
       1
  \end{array} \right)$:
\begin{equation}\label{eq:full_g}
 G= {\bf s}_{\rm i}^\dag \cdot {\bf G} \cdot {\bf s}_{\rm c} = G^{\uparrow\uparrow} +
 G^{\uparrow\downarrow}  + G^{\downarrow\uparrow} + G^{\downarrow \downarrow}.
\end{equation}
Therefore, Fig.~\ref{fig:ucf_spin} also plots the CF properties of the
total conductances $G_X$, $G_Y$, and $G_Z$. For example, $G
\equiv G_Z$ means that the spin part, $|\!\! \uparrow\rangle$ or
$|\!\! \downarrow\rangle$, of the asymptotic scattering
state~(\ref{eq:channel}) is an eigenstate of $\hat{\sigma}_z$,
i.e., fully polarized along the $z$-axis. They should be
contrasted with ${\rm Var}\, G_0$ in Fig.~\ref{fig:ucf_spinless}.
Comparison is facilitated by looking at the reduction factor ${\rm
Var}\,G_0/{\rm Var}\, G$ in Fig.~\ref{fig:reduction}, which also
depends on the direction of magnetization of the ferromagnetic
contacts (i.e., chosen spin-polarization axis). This is due to
the breaking of rotational invariance by the Rashba term (whose
electric field always lies along the $z$-axis). While $G_X$ and
$G_Z$ (characterizing the device having magnetization of the
leads in the plane orthogonal to the direction of virtual Rashba
magnetic field) are similar and exhibit a similar pattern of CF as
a function of disorder strength, the two setups can still be
distinguished by looking at the CF patterns of partial
conductances that are summed in Eq.~(\ref{eq:full_g}) to get full
conductances for unpolarized transport.

Further inspection of the total conductances in
Fig.~\ref{fig:antilocal} shows that below $W \simeq 2$, the
system is in the quasiballistic transport regime ($\ell \protect
\gtrsim L$) where $G_0 > G_X,G_Y,G_Z$ because of the additional 
scattering at the FM-Sm interface which can arise due to band structure 
mismatch~\cite{nikolic_jcmp} induced by the Rashba term. Upon entering the  
diffusive regime antilocalization  $G_0 < G_Z,G_Y,G_Z$ sets in as a standard 
quantum-interference effect (which survives disorder averaging) pertinent to 
all spin-orbit interactions affecting phase-coherent propagation through disorder.~\cite{larkin,iordanskii} 
Thus, strong SO coupling impedes localization effects on both the CF and  
conductances, even at very high disorder. At smaller SO couplings in Fig.~\ref{fig:antilocal}, 
the antilocalization effects are substantially  diminished ($t_{\rm so}^{\rm R}=0.1$) or vanish
$G_0 \approx G_X,G_Y,G_Z$ altogether ($t_{\rm so}^{\rm R}=0.01$).
Nevertheless, the presence of weak SO coupling (i.e., small $t_{\rm so}^{\rm R}/W$) 
is still palpable in the CF plotted in Fig.~\ref{fig:reduction}, and even more so in the relationship
between different partial conductances studied in Sec.~\ref{sec:conductance}.

\section{Spin-resolved partial conductances in disordered
Rashba spin-split electron system} \label{sec:conductance}

The long lifetime of the electron spin orientation has lead to plausible
expectations that the conductance of a two-terminal device
$G^{\uparrow\uparrow}$ with parallel magnetizations in the
ferromagnetic contacts should be higher than
$G^{\uparrow\downarrow}$ when that orientation is
antiparallel.~\cite{datta90,semi1} However, a recent
experiment~\cite{hu} has unraveled  surprising result 
for the disorder-averaged conductance difference $\langle
G^{\uparrow\uparrow}-G^{\uparrow\downarrow} \rangle$: it decreases
upon increasing sample length (InAs containing 2DEG that is
attached to permalloy leads), and eventually turns negative in
the quasiballistic transport regime $\ell \sim L$. Theoretical
modeling~\cite{seba} of the effect of localized spin-flip
interactions, which couple two spin subsystems at a discrete sets
of points within a mesoscopic disordered sample, finds similar
phenomenon under the simplified assumptions for quantum-mechanical
transmissivities (e.g., $T^{\uparrow\downarrow}$) that are
\begin{figure}
\centerline{ \psfig{file=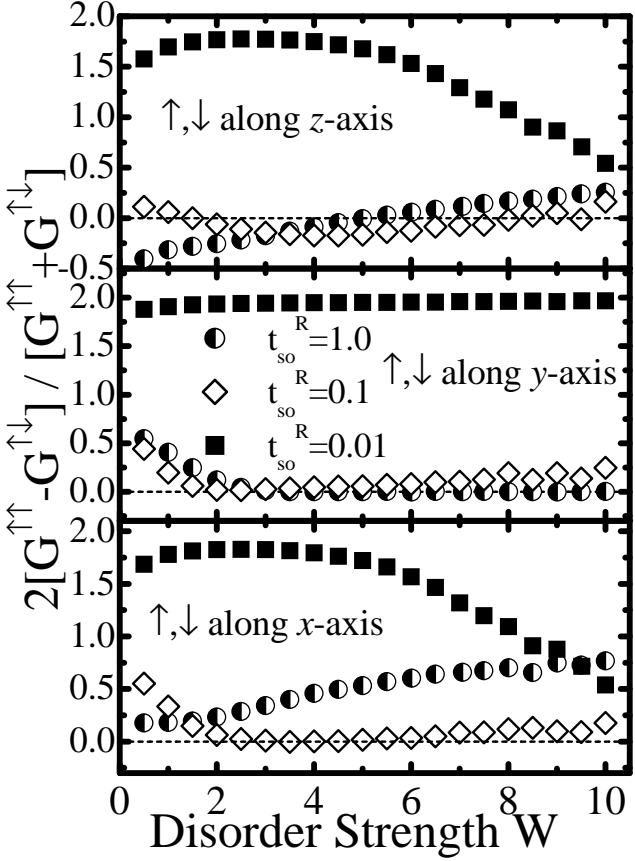,height=4.5in,angle=0} }
 \caption{Disorder-averaged relative conductance difference $\Delta
G^{\uparrow\uparrow}_{\uparrow\downarrow}/G^{\uparrow\uparrow}_{\uparrow\downarrow}$
as functions of the disorder strength (i.e., in different
transport regimes: ballistic, quasiballistic, diffusive, and
localized), Rashba SO coupling, and the direction of magnetization in
the leads with respect to the $z$-axis Rashba electric field..
The injected electrons, into the same set of Rashba spin-split
disordered conductors studied in Fig.~\ref{fig:ucf_spin}, are
spin-up polarized in the direction denoted on each panel.}
\label{fig:surprise}
\end{figure}
\begin{figure}
\centerline{
\psfig{file=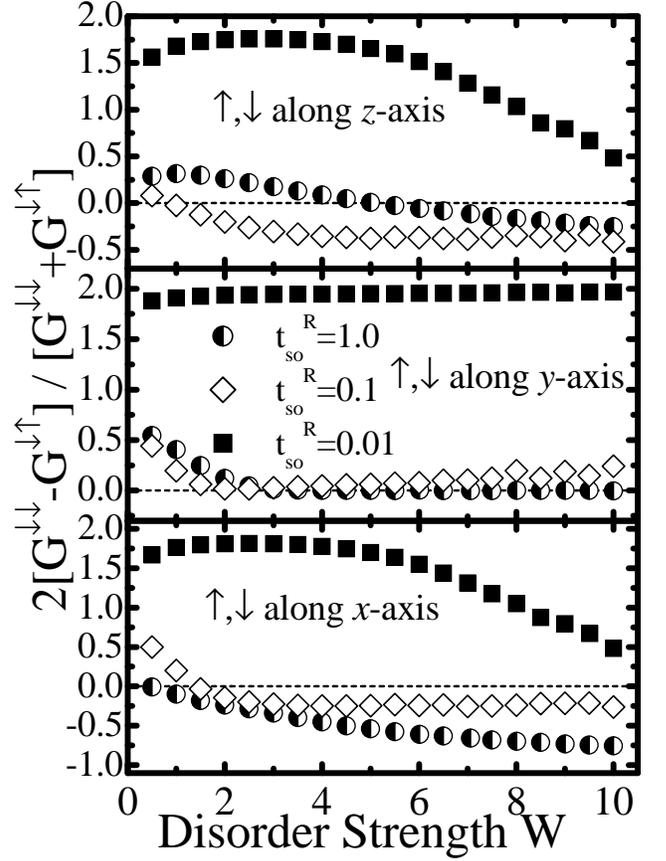,height=4.5in,angle=0} }
 \caption{Disorder-averaged relative conductance difference
 $\Delta G^{\downarrow\downarrow}_{\downarrow\uparrow}/G^{\downarrow\downarrow}_{\downarrow\uparrow}$
as functions of the disorder strength, Rashba SO coupling, and 
the orientation of magnetization in the leads with respect to the
$z$-axis Rashba electric field. The injected electrons, into the
same set of Rashba spin-split disordered conductors studied in
Fig.~\ref{fig:ucf_spin}, are spin-down polarized in the direction
denoted on each panel.} \label{fig:surprise_inverse}
\end{figure}
independent of the spin orientation of the incoming electron. The
results of Sec.~\ref{sec:ucf} point out that such a treatment is not
sufficient in devices with a Rashba SO coupling (as hinted at in
Ref.~\onlinecite{seba}). Therefore, we analyze in Fig.~\ref{fig:surprise} 
the disorder-averaged conductance differences (normalized by respective sums) 
when an $\uparrow$-electron is injected
\begin{equation}\label{eq:c_diff1}
\frac{\Delta
G^{\uparrow\uparrow}_{\uparrow\downarrow}}{G^{\uparrow\uparrow}_{\uparrow\downarrow}}
 =  2 \frac{\langle
G^{\uparrow\uparrow}-G^{\uparrow\downarrow}\rangle}{\langle
G^{\uparrow\uparrow}+G^{\uparrow\downarrow} \rangle},
\end{equation}
while Fig.~\ref{fig:surprise_inverse} plots the corresponding
differences for injection of a spin-down electron
\begin{equation} \label{eq:c_diff2}
 \frac{\Delta
G^{\downarrow\downarrow}_{\downarrow\uparrow}}{G^{\downarrow\downarrow}_{\downarrow\uparrow}}
 =  2 \frac{\langle
G^{\downarrow\downarrow}-G^{\downarrow\uparrow}\rangle}{\langle
G^{\downarrow\downarrow}+G^{\downarrow\uparrow} \rangle},
\end{equation}
using three different cases for the spin-polarizations of incoming
electron with respect to the $z$-axis Rashba electric field. The
set of disorder and SO hopping parameters is the same as in
Sec.~\ref{sec:ucf}. Comparing $\Delta
G^{\uparrow\uparrow}_{\uparrow\downarrow}/G^{\uparrow\uparrow}_{\uparrow\downarrow}$
and $\Delta
G^{\downarrow\downarrow}_{\downarrow\uparrow}/G^{\downarrow\downarrow}_{\downarrow\uparrow}$
shows that the relationship between conductances for parallel and
antiparallel magnetization in the leads is quite intricate: for
the $x$-axis or $z$-axis magnetization, the Rashba spin-split
system is not invariant with respect to the spin-subsystem
interchange; on the other hand, in the case of polarization along
the $y$-axis invariance holds, $\Delta
G^{\uparrow\uparrow}_{\uparrow\downarrow}/G^{\uparrow\uparrow}_{\uparrow\downarrow}
\approx \Delta
G^{\downarrow\downarrow}_{\downarrow\uparrow}/G^{\downarrow\downarrow}_{\downarrow\uparrow}$.
We also find that some of the conductance  differences for the
$x$- and $y$-polarization direction can change sign, even for small SO
coupling ($t_{\rm so}^{\rm R}/W \sim 0.1$). This occurs within
different transport regimes, depending on the polarization of 
the incoming electron, so that negative values of conductance
differences are not confined only to the quasiballistic transport
regime as in Refs.~\onlinecite{seba,hu} (the most similar case
here to their phenomenology is $\uparrow$ polarized, along the
$z$-axis, current injected into the conductor with moderate SO
coupling $t_{\rm so}^{\rm R}=0.1$).

These negative values for conductance differences in 2D mesoscopic
spintronics disordered systems is  a manifestation of a novel
quantum interference effect involving the two spin
components.~\cite{seba} For instance, if an $\uparrow$-electron is
injected into a ballistic sample, the Rashba interaction will induce
spin precession rendering non-zero $G^{\uparrow \downarrow} \neq
0$ because the $\downarrow$-electron could be detected at the drain
even if no spin-down electrons are injected at the source.~\cite{datta90,mireles} 
When scattering off impurities is added to the SO interaction, the injected electron 
pure quantum state, with definite spin and momentum, evolves into another pure 
quantum state that is a linear superposition (as discussed in Sec.~\ref{sec:hamilton}) 
of different and entangled momentum and spin eigenstates. In the intuitive 
(semiclassical) picture of Feynman paths,~\cite{hu} for some device 
configurations shown in Figs.~\ref{fig:surprise} and
\ref{fig:surprise_inverse}, this superposition is equivalent to a
destructive interference taking place along those trajectories
where projection of the electron spin on the magnetization axis
of the source has not yet  changed sign at the drain contact.

\section{Conclusion}\label{sec:conclusion}

Using our efficient real$\otimes$spin space formalism for the
evaluation of partial spin-resolved conductances, expressed in
terms of a Landauer-type formula, we have investigated some of
the standard interference effects in the quantum-coherent
propagation of electrons through two-dimensional disordered
conductor, which is attached to two ferromagnetic contacts
 and is exposed to a Rashba spin-orbit coupling. In 
the limit of a strong Rashba interaction, all computed conductances
exhibit a critical value of $\simeq e^2/h$  in a substantial
interval of large disorder, where $G_0$ for the time-reversal and
spin-rotation invariant system is negligible because of strong
localization effects. The freezing out of the fluctuations of the
total conductances, ${\rm Var}\, G_X$, ${\rm Var}\, G_Y$, and
${\rm Var}\, G_Z$, is even more dramatic, in contrast to the pure
disordered case where the CF decay exponentially fast. The impediment
of the Anderson localization at strong disorder resembles a
weak-antilocalization effect due to SO scattering off
impurities,~\cite{larkin} which is known to generate a
metal-insulator transition in 2D systems (otherwise, they are
Anderson insulators, albeit with an exponentially large $\xi$ at
weak disorder~\cite{verges}). Nevertheless, it has been known
that the Rashba induced antilocalization can be considerably
different~\cite{iordanskii,datta_exp} from the standard
weak-antilocalization expression of Ref.~\onlinecite{larkin}.
While both conductance fluctuations and localization effects are
thought to fall into three universality classes (determined only
by the invariance properties of the Hamiltonian of a conductor
with respect to time-reversal and spin-rotation symmetry
operations~\cite{carlo_rmt}), here we demonstrate how the tracking of
individual spin subsystems (i.e., injection and detection of
spin-polarized electrons) allows one to reveal non-universal 
features specific to the Rashba SO coupling effects on the 
phase-coherent propagation through disorder. For example, one has
to take into account the angle between the spin-polarization of an 
incoming electron and the Rashba electric field. Moreover, the 
relationship between spin-resolved partial conductances unearths
additional quantum interference effects specific to the context of 
spintronics: for particular set of disorder, SO, and spin-polarization parameters, 
one can obtain special superpositions of the two spin components, which, being 
entangled with orbital states of mesoscopic disordered systems, generate  higher 
conductance for the two-terminal device with antiparallel contact magnetization 
than for parallel configuration.

Finally, we emphasize that our findings are {\em numerically
exact} within the single-particle picture of transport, and
therefore non-perturbative in both the strength of the
disorder~\cite{nikolic} and SO interaction~\cite{mireles}. For
example, at large enough  $W$ the concept of mean free path,  
 which is relevant for  semiclassical transport and perturbative 
quantum  interference corrections to it, ceases to  exist~\cite{nikolic} 
($\ell < a \lesssim \lambda_F$). Eventually, the localized phase is encountered 
when $L \gg \xi$.

\begin{acknowledgments}
We thank E. I. Rashba, J. Fabian, D. Frustaglia, and I. \v{Z}uti\'{c} for
enlightening discussions. This work was supported in part by ONR
grant N00014-99-1-0328.
\end{acknowledgments}




\begin{references}
\bibitem{rashba_values} The largest experimentally observed value of the Rashba SO coupling is 
$\alpha_{\rm R} = 3 \cdot 10^{-12}$ eVm in InAs-based heterojunctions.~\cite{nitta} There seems 
to be no fundamental obstacle in making it larger through a proper choice of material, geometry, 
and external gates.
\end{references}
\end{document}